\documentclass[12pt]{article} 
\usepackage{epsfig,axodraw,amssymb}

\def\ltsim{\lower3pt\hbox{$\, \buildrel < \over \sim \, $}}
\def\gtsim{\lower3pt\hbox{$\, \buildrel > \over \sim \, $}}
\def\be{\begin{equation}}
\def\ee{\end{equation}}
\def\ba{\begin{eqnarray}}
\def\ea{\end{eqnarray}}
\def\ga{\mathrel{\raise.3ex\hbox{$>$\kern-.75em\lower1ex\hbox{$\sim$}}}}
\def\la{\mathrel{\raise.3ex\hbox{$<$\kern-.75em\lower1ex\hbox{$\sim$}}}}

\begin{document}

\baselineskip=16pt 
\begin{titlepage}
\rightline{OUTP-00-25P}
\rightline{hep-th/0006030}
\rightline{June 2000}  
\begin{center}

\vspace{0.5cm}

\large {\bf Multi-Brane Worlds and modification of gravity at large scales.}

\vspace*{5mm}
\normalsize

{\bf Ian I. Kogan\footnote{i.kogan@physics.ox.ac.uk}, Stavros
Mouslopoulos\footnote{s.mouslopoulos@physics.ox.ac.uk}, Antonios Papazoglou\footnote{a.papazoglou@physics.ox.ac.uk}\\ 
 and Graham G. Ross \footnote{g.ross@physics.ox.ac.uk}}

\smallskip 
\medskip 
{\it Theoretical Physics, Department of Physics, Oxford University}\\
{\it 1 Keble Road, Oxford, OX1 3NP,  UK}
\smallskip

\vskip0.6in \end{center}
 
\centerline{\large\bf Abstract}

We discuss the implications of multi-brane constructions involving combinations of
positive and negative tension brane and show how anomalously light KK states emerge
when negative tension $''-''$ branes are sandwiched between  $''+''$ branes. We  present a
detailed study of a $''+--+''$ brane assignment which interpolates between  two models that have been 
previously proposed in which gravity is modified at large scales due to the anomalously light states.  We show that it has the peculiar
characteristic that gravity changes from four dimensional (4D) to 5D at large
distances and returns to 4D at even larger scales. We also consider a
crystalline universe which leads to a similar structure for gravity. The problems associated with intermediate negative
tension branes are discussed and a possible resolution suggested.
 
\vspace*{2mm} 

\end{titlepage}

\section{Introduction}

The conjecture that we may live on a brane in a space-time with more than
four dimensions is quite old \cite{80s} but has been the subject of renewed
interest in recent years with the realisation that such structures are
common in string thories. The models of Antoniadis, Arkani-Hamed,
Dimopoulos, Dvali (AADD) \cite{large} and of Randall, Sundrum (RS) \cite
{RS,RS2} have the common characteristic that the Standard Model (SM) fields
are localized on a 3-brane and gravity propagates in all the space-time
dimensions (the ``bulk''). They can provide novel explanations to the Planck
hierarchy problem based either on the size of the bulk volume (AADD), or on
an exponential warp factor that enters the metric (RS). The RS construction
in particular is very attractive because all the parameters of the model may
be of the same magnitude while still generating a very large hierarchy.

The RS model consists of two 3-branes of opposite tension sitting at the
fixed points of an $S^{1}/Z_{2}$ orbifold with $AdS_{5}$ bulk geometry.
Gravity is localized on the brane of positive tension ($^{\prime \prime
}+^{\prime \prime }$ brane) whereas an exponential ``warp'' factor generates
a scale hierarchy on the brane of negative tension ($^{\prime \prime
}-^{\prime \prime }$ brane). There are two possible physical interpretations
of this model. Placing the Standard Model (SM) fields on the $^{\prime
\prime }-^{\prime \prime }$ brane \cite{RS} provides an explanation of the
Planck hierarchy problem. Placing the SM fields on the $^{\prime \prime
}+^{\prime \prime }$ brane \cite{RS2} allows the $^{\prime \prime }-^{\prime
\prime }$ brane to be put at infinity, {\textit{i.e.}} decompactify the
orbifold. In this case, \ however, gravity is localised on the $^{\prime
\prime }+^{\prime \prime }$ brane so the world still appears four dimensional
(see also \cite{gog}). As far as the the first possibility is concerned,
there has been a lot of discussion \cite{cosm} whether we can have
acceptable theory of gravity and cosmology if we live on a $^{\prime \prime
}-^{\prime \prime }$ brane. To avoid this problem   Lykken and Randall 
(LR) \cite{lykken} proposed a $^{\prime \prime }++-^{\prime \prime }$ brane
configuration in which the SM fields reside on the intermediate positive
tension brane (which has vanishingly small tension) and which still has a
warp factor explanation of the mass hierarchy.

In reference \cite{Oxford} we considered an alternative $^{\prime \prime
}+-+^{\prime \prime }$configuration, again with the SM fields on a positive
tension brane and a warp factor. This model (KMPRS) has a novel
phenomenology \cite{gang2} quite distinct from the RS and LR models. This is
mainly due to the intermediate $^{\prime \prime }-^{\prime \prime }$ brane
leading to an anomously light first Kaluza-Klein (KK) as discussed below.
This feature gives rise to the exotic possibility of ``bigravity'', in which
the first KK state has a Compton wavelength of the order of $10^{26}$cm
corresponding to $1\%$ of the size of the observable universe, while, at the
same time, all the remaining states of the KK tower have Compton wavelengths
below $1$mm. In such a scenario the gravitational attraction as we feel it
is the net effect of the exchange of the ordinary 4D graviton {\textit{and}}
the ultralight KK state. This leads to the prediction of modifications of
gravity not only at the millimeter scale (due to the higher KK modes) but
also the ultralarge scale (of ${\mathcal{O}}(10^{26}{\rm{cm}})$!) Indeed, gravity becomes
weaker at ultralarge distances due to the Yukawa suppression of the first KK
mode.

Independently and again in the context of the RS scenario, Gregory, Rubakov,
Sibiryakov (GRS) have recently suggested \cite{GRS} a construction in which
gravity is also modified at both small and at ultralarge scales. The GRS
model modifies the decompactified RS model\cite{RS2} by adding a $^{\prime
\prime }-^{\prime \prime }$ brane of half the tension of the $^{\prime
\prime }+^{\prime \prime }$ brane and requiring flat space to the right of
the new $^{\prime \prime }-^{\prime \prime }$ brane. This model does not
have a normalizable 4D graviton but generates 4D gravity at intermediate
distances due to a resonance-like behaviour of the wavefunctions of the KK
states continuum. Gravity in this picture appears to be ``quasi-localized''
and for ultra large scales becomes five dimensional. In this scenario
gravity is also modified at short scales (but this time much shorter than
the millimeter scale).

Although the KMPRS and GRS models look quite different, they share the key
element of freely moving $^{\prime \prime }-^{\prime \prime }$ branes which
generate the tunneling effects responsible for the anomalously light states.
It was shown in \cite{KR} that these two models are the limiting cases of a
more general $^{\prime \prime }+--+^{\prime \prime }$ multi-brane model that
interpolates between the ``bigravity'' KMPRS model and the
``quasi-localized'' gravity GRS model. This model can be readily derived
from the KMPRS model by cutting in half the intermediate $^{\prime \prime
}-^{\prime \prime }$ brane and considering flat space between the two new $%
^{\prime \prime }-^{\prime \prime }$ branes. Moreover, in \cite{KR} it was
conjectured that every construction having $^{\prime \prime }-^{\prime
\prime }$ branes between $^{\prime \prime }+^{\prime \prime }$ branes is
capable of giving a ``multigravity'' scenario where gravity is generated not
only by the 4D graviton but also by a number (finite or infinite) of KK
states. The crystal universe considered by \cite{kaloper} (see also \cite
{nam}) is an interesting example{\footnote{%
We disagree with the statements made in \cite{kaloper} about the absence of
a zero mode band.}}.
 Let us  mention here that recently Gorsky and Selivanov \cite{GS} 
discussed models where the effect of  negative tension branes   
  can be mimicked by a constant four form field and the presence of the
three brane junctions. It will be interesting to see if in these models
 one has multigravity.

The purpose of this paper is threefold. We first we will consider the $%
^{\prime \prime }++-$ $^{\prime \prime }$ LR construction but with the
intermediate brane carrying non-negligible tension. We show that this model
does not have an anomously light KK state and hence cannot generate a
``multigravity '' scenario. Analysis of the reason for this suggests that $%
^{\prime \prime }+^{\prime \prime }$ branes not separated by $^{\prime
\prime }-^{\prime \prime }$ branes do not lead to anomalously light modes.
Secondly, we examine in detail the $^{\prime \prime }+--+^{\prime \prime }$
``multigravity'' model and show exactly how it interpolates between the
KMPRS and the GRS model. Gravity in this case has the quite peculiar
characteristic that it changes from 4D to 5D at ultra-large scales and then
again to 4D at even larger scales. We also analyse the infinite crystalline
universe and show how the band structure of the KK modes gives rise to
``multigravity''. Finally we discuss the role of the radion and the question
whether these theories give rise to viable theories.

\section{The three-brane $^{\prime \prime }++-^{\prime \prime }$ Model}

\begin{figure}[h]
\begin{center}
\begin{picture}(300,125)(0,50)

\SetWidth{1.5}

\SetOffset(0,10)

\BCirc(150,100){60}
\DashLine(90,100)(210,100){3}

\GCirc(90,100){7}{0.9}
\GCirc(183,148){7}{0.9}
\GCirc(183,52){7}{0.9}

\GCirc(210,100){7}{0.2}

\Text(70,100)[]{$+$}
\Text(230,100)[]{$-$}
\Text(190,162)[]{$+$}
\Text(190,38)[]{$+$}
\Text(170,140)[]{$L_1$}
\Text(193,112)[]{$L_2$}
\Text(158,60)[]{$-L_1$}

\Text(130,120)[]{$Z_2$}

\LongArrowArc(150,100)(68,4,52)
\LongArrowArcn(150,100)(68,52,4)
\Text(220,135)[l]{$x=k_2(L_2-L_1)$}

\SetWidth{2}
\LongArrow(150,100)(150,115)
\LongArrow(150,100)(150,85)

\end{picture}
\end{center}
\caption{The brane locations in the three-brane $^{\prime \prime
}++-^{\prime \prime }$ model. The bulk curvature between the $^{\prime
\prime }+^{\prime \prime }$ branes is $k_{1}$ and between the $^{\prime
\prime }+^{\prime \prime }$ and $^{\prime \prime }-^{\prime \prime }$ brane
is $k_{2}$.}
\label{++-}
\end{figure}
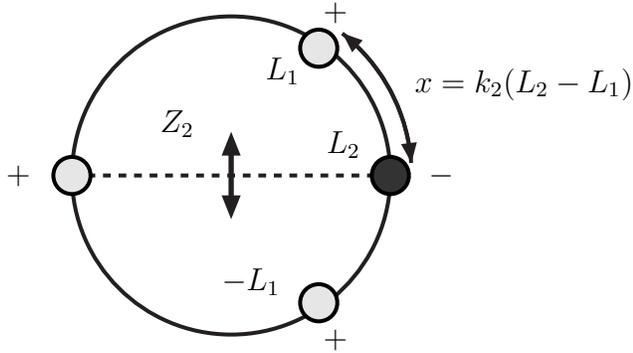

The $^{\prime \prime }++-^{\prime \prime }$ model (see Fig.\ref{++-})
consists of three parallel 3-branes in $AdS_{5}$ spacetime with orbifold
topology, two of which are located at the orbifold fixed points $L_{0}=0$
and $L_{2}$ while the third one is moving at distance $L_{1}$ in between. In
order to get 4D flat space with this configuration, it turns out that the $%
AdS_{5}$ space must have different cosmological constants $\Lambda _{1}$ and 
$\Lambda _{2}$ between the first - second and the second - third brane
respectively with $|\Lambda _{2}|>|\Lambda _{1}|$ (see \cite{many} for
constructions of different bulk cosmological constants). The action of the
above setup (if we neglect the matter contribution on the branes in order to
find a suitable vacuum solution) is given by: 
\begin{eqnarray}
S &=&\int d^{4}x\int_{-L_{2}}^{L_{2}}dy\sqrt{-G}[-\Lambda (y)+2M^{3}R%
]-\sum_{i}\int_{y=y_{i}}d^{4}xV_{i}\sqrt{-\hat{G}^{(i)}}  \label{action} \\
\mathrm{with}~~~\Lambda (y) &=&\left\{ 
\begin{array}{cl}
{\Lambda _{1}} & ,y\in \lbrack 0,L_{1}] \\ 
\Lambda _{2} & ,y\in \lbrack L_{1},L_{2}]
\end{array}
~~~~~~\right. \   \nonumber
\end{eqnarray}
where $\hat{G}_{\mu \nu }^{(i)}$ is the induced metric on the branes, $V_{i}$
are their tensions and $M$ the 5D fundamental scale. We consider as in \cite
{Oxford} the vacuum metric ansatz that respects 4D Poincar\'{e} invariance: 
\begin{equation}
ds^{2}=e^{-2\sigma (y)}\eta _{\mu \nu }dx^{\mu }dx^{\nu }+dy^{2}
\label{ansatz}
\end{equation}

The above ansatz substituted in the Einstein equations requires that $\sigma
(y)$ satisfies the differential equations: 
\begin{equation}
\sigma ^{\prime \prime }=\sum_{i}\frac{V_{i}}{12M^{3}}\delta (y-L_{i})~~~%
\mathrm{and}~~~\left( \sigma ^{\prime }\right) ^{2}=\left\{ 
\begin{array}{cl}
{k_{1}^{2}} & ,y\in \lbrack 0,L_{1}] \\ 
k_{2}^{2} & ,y\in \lbrack L_{1},L_{2}]
\end{array}
\right. \ 
\end{equation}
where $k_{1}=\sqrt{\frac{-\Lambda _{1}}{24M^{3}}}$ and $k_{2}=\sqrt{\frac{%
-\Lambda _{2}}{24M^{3}}}$ are effectively the bulk curvatures in the two
regions between the branes. The solution of these equations for $y>0$ it is
given by: 
\begin{equation}
\sigma (y)=\left\{ 
\begin{array}{cl}
{k_{1}y} & ,y\in \lbrack 0,L_{1}] \\ 
{k_{2}(y-L_{1})+k_{1}L_{1}} & ,y\in \lbrack L_{1},L_{2}]
\end{array}
\right. \ 
\end{equation}
Einstein's equations impose the following conditions on the brane tensions : 
\begin{equation}
V_{0}=24M^{3}k_{1},~~~V_{1}=24M^{3}\frac{k_{2}-k_{1}}{2}%
,~~~V_{2}=-24M^{3}k_{2}
\end{equation}

If we consider massless fluctuations of this vacuum metric and then
integrate over the 5-th dimension, we find the 4D Planck mass is given by: 
\begin{equation}
M_{\mathrm{Pl}}^2={M^3}\left[\frac{1}{k_1}\left(1-e^{-2k_1L_1}\right)+\frac{1%
}{k_2}e^{2(k_2-k_1)L_1}\left(e^{-2k_{2}L_1}-e^{-2k_{2}L_{2}}\right)\right]
\end{equation}

The above formula tells us that for large enough $kL_{1}$ and $kL_{2}$ the
four mass scales $M_{\mathrm{Pl}}$, $M$, $k_{1}$ and $k_{2}$ can be taken to
be of the same order. Thus we take $k_{1},k_{2}\sim {\mathcal{O}}(M)$ in
order not to introduce a new hierarchy, with the additional restriction $%
k_{1}<k_{2}<M$ so that the bulk curvature is small compared to the 5D Planck
scale and we can trust the solution. Furthermore, if we put matter on the
second brane all the physical masses $m$ on it will be related to the mass
parameters $m_{0}$ of the fundamental 5D theory by the conformal ``warp''
factor 
\begin{equation}
m=e^{-\sigma \left( L_{1}\right) }m_{0}=e^{-k_{1}L_{1}}m_{0}
\end{equation}

This allows us to put the SM states on the intermediate $^{\prime \prime
}+^{\prime \prime }$ brane, solving the Planck hierarchy problem by choosing 
$e^{-kL_{1}}$ to be of $\mathcal{O}$$\left( 10^{-15}\right) $, \textit{i.e} $%
L_{1}\approx 35k^{-1}$.

The KK spectrum can be as usual found by considering the linear ``massive''
fluctuations of the metric. We use the following convention\footnote{%
Here we have ignored the radion modes that could be used to stabilize the
brane positions $L_{1}$ and $L_{2}$. For discussion and possible
stabilization mechanisms see \cite{stab}}: 
\begin{equation}
ds^{2}=\left[ e^{-2\sigma (y)}\eta _{\mu \nu }+\frac{2}{M^{3/2}}h_{\mu \nu
}(x,y)\right] dx^{\mu }dx^{\nu }+dy^{2}  \label{fluct}
\end{equation}

We expand the field $h_{\mu \nu }(x,y)$ in terms of the 4D graviton zero
mode and the KK states plane waves $h_{\mu \nu }^{(n)}(x)$: 
\begin{equation}
h_{\mu \nu }(x,y)=\sum_{n=0}^{\infty }h_{\mu \nu }^{(n)}(x)\Psi ^{(n)}(y)
\end{equation}
where $\left( \partial _{\kappa }\partial ^{\kappa }-m_{n}^{2}\right) h_{\mu
\nu }^{(n)}=0$ after fixing the gauge as $\partial ^{\alpha }h_{\alpha \beta
}^{(n)}=h_{\phantom{-}\alpha }^{(n)\alpha }=0$. The wavefunction $\Psi
^{(n)}(y)$ obeys a second order differential equation and carries all the
information about the effect of the non-factorizable geometry on the
graviton and the KK states. After a change of coordinates and a redefinition
of the wavefunction the problem reduces to the solution of an ordinary
Schr\"{o}dinger equation: 
\begin{equation}
\left\{ -\frac{1}{2}\partial _{z}^{2}+V(z)\right\} \hat{\Psi}^{(n)}(z)=\frac{%
m_{n}^{2}}{2}\hat{\Psi}^{(n)}(z)  \label{sch}
\end{equation}
where the potential $V(z)$ for $z>0$ has the form: 
\begin{eqnarray}
\hspace*{0.5cm}V(z) &=&\frac{15}{8[g(z)]^{2}}\left[ k_{1}^{2}(\theta
(z)-\theta (z-z_{1}))+k_{2}^{2}(\theta (z-z_{1})-\theta (z-z_{2}))\right] 
\nonumber \\
&&-\frac{3}{2g(z)}\left[ k_{1}\delta (z)+\frac{(k_{2}-k_{1})}{2}\delta
(z-z_{1})-k_{2}\delta (z-z_{2})\right]  \label{pot}
\end{eqnarray}

The new coordinates and wavefunction in the above equations are defined by: 
\begin{equation}
\renewcommand{\arraystretch}{1.5}z\equiv \left\{ 
\begin{array}{cl}
\frac{e^{k_{1}y}-1}{k_{1}} & ,y\in \lbrack 0,L_{1}] \\ 
\frac{e^{k_{2}(y-L_{1})+k_{1}L_{1}}}{k_{2}}+\frac{e^{k_{1}L_{1}}-1}{k_{1}}-%
\frac{e^{k_{1}L_{1}}}{k_{2}} & ,y\in \lbrack L_{1},L_{2}]
\end{array}
\right. \ 
\end{equation}
\begin{equation}
\hat{\Psi}^{(n)}(z)\equiv \Psi ^{(n)}(y)e^{\sigma /2}
\end{equation}
with the symmetric $z$ for $y<0$ and the function $g(z)$ as: 
\begin{equation}
g(z)=\left\{ 
\begin{array}{cl}
{k_{1}z+1} & ,z\in \lbrack 0,z_{1}] \\ 
{k_{2}(z-z_{1})+k_{1}z_{1}+1} & ,z\in \lbrack z_{1},z_{2}]
\end{array}
\right. \ 
\end{equation}
where $z_{1}=z(L_{1})$ and $z_{2}=z(L_{2})$. The change of coordinates has
been chosen so that there are no first derivative terms in the differential
equation. Furthermore, in this coordinate system it can be easily seen that
the vacuum metric takes the conformaly flat form: 
\begin{equation}
ds^{2}=\frac{1}{g(z)^{2}}\left( \eta _{\mu \nu }dx^{\mu }dx^{\nu
}+dz^{2}\right)
\end{equation}

The potential (\ref{pot}) always gives rise to a massless graviton zero mode
which reflects the fact that Lorentz invariance is preserved in 4D
spacetime. Its wavefunction is given by: 
\begin{equation}
\hat{\Psi}^{(0)}=\frac{A}{[g(z)]^{3/2}}  \label{zerowave}
\end{equation}
with normalization factor $A$ determined by the requirement $\displaystyle{%
\int_{-z_2}^{\phantom{-}z_2} dz\left[\hat{\Psi}^{(0)}(z)\right]^2=1}$,
chosen so that we get the standard form of the Fierz-Pauli Lagrangian (the
same holds for the normalization of all the other states).

For the massive KK modes the solution is given in terms of Bessel functions.
For $y$ lying in the regions $\mathbf{A}\equiv\left[0,L_1\right]$ and $%
\mathbf{B}\equiv\left[L_1,L_2\right]$, we have: 
\begin{equation}
\hat{\Psi}^{(n)}\left\{ 
\begin{array}{cc}
\mathbf{A} &  \\ 
\mathbf{B} & 
\end{array}
\right\}=N_n \renewcommand{\arraystretch}{2} \left\{ 
\begin{array}{cc}
\sqrt{\frac{g(z)}{k_{1}}}\left[\phantom{A_1}Y_2\left(\frac{m_n}{k_{1}}%
g(z)\right)+A_{\phantom{2}}J_2\left(\frac{m_n}{k_{1}}g(z)\right)\right] & 
\\ 
\sqrt{\frac{g(z)}{k_{2}}}\left[B_1Y_2\left(\frac{m_n}{k_{2}}%
g(z)\right)+B_2J_2\left(\frac{m_n}{k_{2}}g(z)\right)\right] & 
\end{array}
\right\}  \label{wave}
\end{equation}

The boundary conditions (one for the continuity of the wavefunction at $%
z_{1} $ and three for the discontinuity of its first derivative at $0$, $%
z_{1}$, $z_{2}$) result in a $4\times 4$ homogeneous linear system which, in
order to have a non-trivial solution, should have a vanishing determinant: 
\begin{equation}
\renewcommand{\arraystretch}{1.5}\left| 
\begin{array}{cccc}
Y_{1}\left( \frac{m}{k_{1}}\right) & J_{1}\left( \frac{m}{k_{1}}\right) & %
\phantom{-}0 & \phantom{-}0 \\ 
0 & 0 & \phantom{---}Y_{1}\left( \frac{m}{k_{2}}g(z_{2})\right) & %
\phantom{---}J_{1}\left( \frac{m}{k_{2}}g(z_{2})\right) \\ 
Y_{1}\left( \frac{m}{k_{1}}g(z_{1})\right) & J_{1}\left( \frac{m}{k_{1}}%
g(z_{1})\right) & -\sqrt{\frac{k_{1}}{k_{2}}}Y_{1}\left( \frac{m}{k_{2}}%
g(z_{1})\right) & -\sqrt{\frac{k_{1}}{k_{2}}}J_{1}\left( \frac{m}{k_{2}}%
g(z_{1})\right) \\ 
Y_{2}\left( \frac{m}{k_{1}}g(z_{1})\right) & J_{2}\left( \frac{m}{k_{1}}%
g(z_{1})\right) & -\sqrt{\frac{k_{1}}{k_{2}}}Y_{2}\left( \frac{m}{k_{2}}%
g(z_{1})\right) & -\sqrt{\frac{k_{1}}{k_{2}}}J_{2}\left( \frac{m}{k_{2}}%
g(z_{1})\right)
\end{array}
\right| =0  \label{det}
\end{equation}
(The subscript $n$ on the masses $m_{n}$ has been suppressed.)

\subsection{{\protect\Large \textbf{Masses and Couplings}}}

The above quantization condition determines the mass spectrum of the model.
The parameters we have are $k_{1}$, $k_{2}$, $L_{1}$, $L_{2}$ with the
restriction $k_{1}<k_{2}<M$ and ${k_{1}}\sim {k_{2}}$ so that we don't
introduce a new hierarchy. It is more convenient to introduce the parameters 
$x=k_{2}(L_{2}-L_{1})$, $w=e^{-k_{1}L_{1}}$ and work instead with the set $%
k_{1}$, $k_{2}$, $x$, $w$. From now on we will assume that $w\ll 1$ ($w\sim {%
\mathcal{O}}\left( 10^{-15}\right) $ as is required if the model is to
provide an explanation of the hierarchy problem).

For the region $x\gtrsim 1$ it is straightforward to show analytically that
all the masses of the KK tower scale in the same way as $x$ is varied: 
\begin{equation}
m_{n}=\xi _{n}k_{2}we^{-x}  \label{big}
\end{equation}
where $\xi _{n}$ is the $n$-th root of $J_{1}(x)$. This should be compared
with the KMPRS $^{\prime \prime }+-+^{\prime \prime }$model in which $%
m_{1}\propto kwe^{-2x}$ and $m_{n+1}\propto kwe^{-x}$, where $x$ is the
separation between the $^{\prime \prime }-^{\prime \prime }$ and the second $%
^{\prime \prime }+^{\prime \prime }$ brane. This significant difference can
be explained by the fact that in the $^{\prime \prime }++-^{\prime \prime }$%
case the negative tension brane creates a potential barrier between the two
attractive potentials created by the positive tension branes. As a result
the wave function in the region of the $^{\prime \prime }-^{\prime
\prime }$ brane is small due to the tunneling effect. The two attractive potentials
support two bound states, one the graviton and the other the first KK mode.
The mass difference between the two is determined by the wave function in
the neighbourhood of the $^{\prime \prime }-^{\prime \prime }$ brane and is
thus very small. On the other hand the wave function between the two $%
^{\prime \prime }+^{\prime \prime }$ branes in the $^{\prime \prime
}++-^{\prime \prime }$ configuration is not suppressed by the need to tunnel
and hence the mass difference between the zero mode and the first KK mode is
also not suppressed. This has as result the two $^{\prime \prime }+^{\prime
\prime }$ branes behave approximately as one. This becomes even more clear
when $x\gg 1$ where the model resembles the $^{\prime \prime }+-^{\prime
\prime }$ RS construction. Indeed, in this limit the mass spectrum becomes $%
m_{n}=\xi _{n}k_{2}e^{-k_{2}L_{2}}$ which is exactly that of the $^{\prime
\prime }+-^{\prime \prime }$ RS model with orbifold size $L_{2}$ and bulk
curvature $k_{2}$.

In the region $x\lesssim 1$ the relation of eq(\ref{big}) breaks down. As
reduce $x$ the second $^{\prime \prime }+^{\prime \prime }$ comes closer and
closer to the $^{\prime \prime }-^{\prime \prime }$ brane and in the limit $%
x\rightarrow 0$ \mbox{({\textit {i.e.}}
$L_{2}=L_{1}$)} the combined brane system behaves as a single $^{\prime
\prime }-^{\prime \prime }$ brane, reducing to the $^{\prime \prime
}+-^{\prime \prime }$ RS model. In this limit the spectrum is given by: 
\begin{equation}
m_{n}=\xi _{n}k_{1}w  \label{small}
\end{equation}
which is just the one of the $^{\prime \prime }+-^{\prime \prime }$ RS
model. In the region $0\leqslant x\lesssim 1$ the mass spectrum interpolates
between the relations (\ref{big}) and (\ref{small}).

The fact that there is nothing special about the first KK mode is true also
for its coupling on the second $^{\prime \prime }+^{\prime \prime }$ brane.
The interaction of the KK states on the second $^{\prime \prime }+^{\prime
\prime }$ brane is given by: 
\begin{equation}
{\mathcal{L}}_{int}=\sum_{n\geq 0}a_{n}h_{\mu \nu }^{(n)}(x)T_{\mu \nu
}(x)~~,~~\mathrm{with}~~a_{n}=\left[ \frac{g(z_{1})}{M}\right] ^{3/2}\hat{%
\Psi}^{(n)}(z_{1})  \label{coupl}
\end{equation}

In the RS limit ($x=0$) all the states of the KK tower have equal coupling
given by: 
\begin{equation}
a_{n}=\frac{1}{wM_{\mathrm{Pl}}}
\end{equation}

As we increase $x$, the lower a state is in the tower, the more strongly it
couples, {\textit{i.e.}} $a_{1}>a_{2}>a_{3}>\cdots $ (with $a_{1}<(wM_{%
\mathrm{Pl}})^{-1}$) and tends to a constant value for high enough levels.
At some point this behaviour changes, the levels cross and for $x\gtrsim 1$
the situation is reversed and the lower a state is in the tower, the more
weakly it couples, {\textit{i.e.}} $a_{1}<a_{2}<a_{3}<\cdots $. In this
region it is possible to obtain a simple analytical expression for the
couplings: 
\begin{equation}
a_{n}=\frac{8\xi _{n}^{2}}{J_{2}\left( \xi _{n}\right) }\left( \frac{k_{2}}{%
k_{1}}\right) ^{3/2}\frac{1}{wM_{Pl}}e^{-3x}
\end{equation}
Here, we also see that the first KK state scales in exactly the same way as
the remaining states in the tower with respect to $x$, a behaviour quite
different to that in the KMPRS model in which the coupling is $x$%
-independent. Furthermore, the coupling falls as ${e^{-3x}}$, {\textit{i.e.}}
much faster than $e^{-x}$ as one would naively expect. This can be explained
by looking at the origin of the $x$-dependence of the wavefunction. For
increasing $x$ the normalization volume coming from the region between $%
L_{1} $ and $L_{2}$ dominates and the normalization constant in (\ref{wave})
scales as $N_{n}\propto e^{-3x}$. This rapid decrease is not compensated by
the increase of the value of the remaining wavefunction (which from (\ref
{wave}) is approximately constant). Thus, although the two $^{\prime \prime
}+^{\prime \prime }$ branes in the large $x$ limit behave as one as far as
the mass spectrum is concerned, their separation actually makes the coupling
of the KK modes very different.

\section{The four-brane $^{\prime \prime }+--+^{\prime \prime }$ Model}

In this section we discuss the four-brane model in order to clarify the
relation between the ``bigravity'' KMPRS model \cite{Oxford} with the GRS
model\cite{GRS}. In particular we wish to explore and compare the
modification of gravity at large scales predicted by each model.

In the case of the KMPRS $^{\prime \prime }+-+^{\prime \prime }$ model, in
the limit of very large $x$, gravity results from the net effect of both the
massless graviton and the ultralight first KK state. The modifications of
gravity at very large distances come from the fact that the Yukawa type
suppression of the gravitational potential coming from the KK state turns on
at the Compton wavelength of the state. On the other hand, the GRS model has
a continuous spectrum with no normalizable zero mode. However, the values of
the KK states wavefunctions on the $^{\prime \prime }+^{\prime \prime }$
brane have a ``resonance''-like behaviour \cite{CEH} which give rise to 4D
gravity at distances smaller than the Compton wavelength of its width.
Beyond this scale gravity becomes five-dimensional.

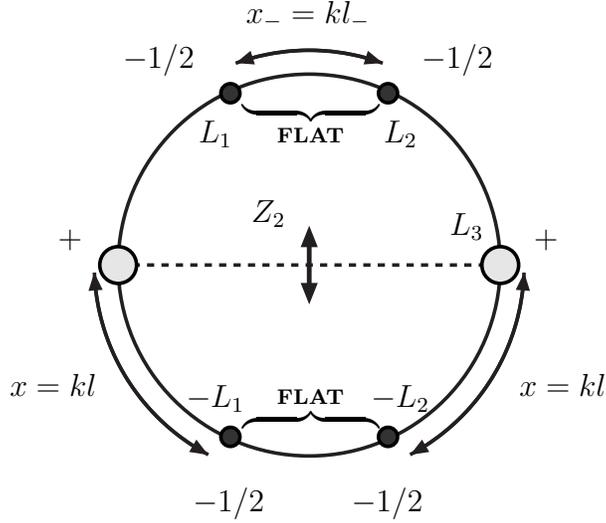
\begin{figure}[t]
\begin{center}
\begin{picture}(300,160)(0,50)

\SetScale{0.9}
\SetOffset(20,40)
\SetWidth{1.5}

\BCirc(150,100){80}
\DashLine(70,100)(230,100){3}

\GCirc(70,100){7.7}{0.9}
\GCirc(230,100){7.7}{0.9}

\GCirc(183,172){4}{0.2}
\GCirc(183,28){4}{0.2}
\GCirc(117,172){4}{0.2}
\GCirc(117,28){4}{0.2}

\LongArrowArc(150,100)(90,298,358)
\LongArrowArcn(150,100)(90,358,298)
\LongArrowArc(150,100)(90,182,242)
\LongArrowArcn(150,100)(90,242,182)

\LongArrowArc(150,100)(90,70,110)
\LongArrowArcn(150,100)(90,110,70)

\Text(45,100)[]{$+$}
\Text(225,100)[]{$+$}
\Text(178,168)[l]{$-1/2$}
\Text(92,168)[r]{$-1/2$}
\Text(165,0)[c]{$-1/2$}
\Text(105,0)[c]{$-1/2$}

\Text(120,110)[]{$Z_2$}

\Text(215,45)[l]{$x=kl$}
\Text(55,45)[r]{$x=kl$}
\Text(135,185)[c]{$x_-=kl_-$}

\Text(100,140)[c]{$L_1$}
\Text(100,40)[c]{$-L_1$}
\Text(170,140)[c]{$L_2$}
\Text(170,40)[c]{$-L_2$}
\Text(195,105)[c]{$L_3$}

\Text(136,149)[c]{$\underbrace{\phantom{abcdefghi}}_{{\Large{\bf FLAT}}}$}
\Text(136,31)[c]{$\overbrace{\phantom{abcdefghi}}^{{\LARGE{\bf FLAT}}}$}

\SetWidth{2}
\LongArrow(150,100)(150,115)
\LongArrow(150,100)(150,85)

\end{picture}
\end{center}
\caption{$^{\prime\prime}+--+^{\prime\prime}$ configuration with scale
equivalent $^{\prime\prime}+^{\prime\prime}$ branes. The distance between
the $^{\prime\prime}+^{\prime\prime}$ and $^{\prime\prime}-1/2^{\prime%
\prime} $ branes is $l=L_1=L_3-L_2$ while the distance between the $%
^{\prime\prime}-1/2^{\prime\prime}$ branes is $l_-=L_2-L_1$. The curvature
of the bulk between the $^{\prime\prime}+^{\prime\prime}$ and $%
^{\prime\prime}-1/2^{\prime\prime}$ branes is $k$.}
\label{multi}
\end{figure}

The four-brane GRS configuration can be obtained from the KMPRS model by
``cutting'' the $^{\prime \prime }-^{\prime \prime }$ brane in half, {%
\textit{i.e.}} instead of having one $^{\prime \prime }-^{\prime \prime }$
brane one can take two $^{\prime \prime }-^{\prime \prime }$ branes of half
the tension of the original one ($^{\prime \prime }-1/2^{\prime \prime }$
branes), having flat spacetime between them (see Fig.(\ref{multi}). Finally
if the second $^{\prime \prime }+^{\prime \prime }$ brane is taken to
infinity together with one of the $^{\prime \prime }-1/2^{\prime \prime }$
branes we shall get precisely the GRS picture.

Let us discuss the four-brane $^{\prime \prime }+--+^{\prime \prime }$ model
in more detail. It consists of 5D spacetime with orbifold topology with four
parallel 3-branes located at $L_{0}=0$, $L_{1}$, $L_{2}$ and $L_{3}$, where $%
L_{0}$ and $L_{3}$ are the orbifold fixed points (see Fig.(\ref{multi})). The
bulk cosmological constant $\Lambda $ is negative ({\textit{i.e.}} $AdS_{5}$
spacetime) between the branes with opposite tension and zero ({\textit{i.e.}}
flat spacetime) between the two $^{\prime \prime }-1/2^{\prime \prime }$
branes. The model has four parameters namely $L_{1}$, $L_{2}$ and $L_{3}$
and $\Lambda $. For our present purposes we consider the symmetric
configuration, leaving 3 parameters, $l$, $l_{-}$ and $\Lambda $ where
$l\equiv L_1=L_3-L_2$ and $l_{-}\equiv L_{2}-L_{1}$. In the absence of matter the model is described by
eq(\ref{action}) with 
\begin{equation}
\Lambda (y)=\left\{ 
\begin{array}{cl}
{0} & ,y\in \lbrack L_{1},L_{2}] \\ 
{\Lambda}  & ,y\in \lbrack 0,L_{1}]\bigcup [L_{2},L_{3}]
\end{array}
\right.
\end{equation}

By considering the ansatz eq(\ref{ansatz}) the ``warp'' function $\sigma (y)$
must satisfy: 
\begin{equation}
\sigma ^{\prime \prime }=\sum_{i}\frac{V_{i}}{12M^{3}}\delta (y-L_{i})~~~%
\mathrm{and}~~~\left( \sigma ^{\prime }\right) ^{2}=\left\{ 
\begin{array}{cl}
{0} & ,y\in \lbrack L_{1},L_{2}] \\ 
k^{2} & ,y\in \lbrack 0,L_{1}] \bigcup [L_{2},L_{3}]
\end{array}
\right. \ 
\end{equation}
where $k=\sqrt{\frac{-\Lambda }{24M^{3}}}$ is a measure of the bulk
curvature and we take $V_{0}=V_{3}=-2V_{1}=-2V_{2} \equiv
V$. The solution for $y>0$ is: 
\begin{equation}
\sigma (y)=\left\{ 
\begin{array}{cl}
{ky} & ,y\in \lbrack 0,L_{1}] \\ 
{kL_{1}} & ,y\in \lbrack L_{1},L_{2}] \\ 
{kL_{1}+k(L_{2}-y)} & ,y\in \lbrack L_{2},L_{3}]
\end{array}
\right. \ 
\end{equation}
Furthermore, 4D Poincare invariance requires the fine tuned relation: 
\begin{equation}
V=-\frac{\Lambda }{k}
\end{equation}

In order to determine the mass spectrum and the couplings of the KK modes we
consider linear ``massive'' metric fluctuations as in eq(\ref{fluct}).
Following the same procedure we find that the function $\hat{\Psi}^{(n)}(z)$
obeys a Schr\"{o}dinger-like equation with potential $V(z)$ of the form: 
\begin{eqnarray}
\hspace*{0.5cm}V(z) &=&\frac{15k^{2}}{8[g(z)]^{2}}\left[ \theta (z)-\theta
(z-z_{1})+\theta (z-z_{2})-\theta (z-z_{3})\right]  \nonumber \\
&&-\frac{3k}{2g(z)}\left[ \delta (z)-\frac{1}{2}\delta (z-z_{1})-\frac{1}{2}%
\delta (z-z_{2})+\delta (z-z_{3})\right]  \label{pot2}
\end{eqnarray}
The conformal coordinates now are given by: 
\begin{equation}
\renewcommand{\arraystretch}{1.5}z\equiv \left\{ 
\begin{array}{cl}
\frac{e^{ky}-1}{k} & ,y\in \lbrack 0,L_{1}] \\ 
\ (y-l)e^{kl}+\frac{e^{kl}-1}{k} & ,y\in \lbrack L_{1},L_{2}] \\ 
\ -\frac{1}{k}e^{2kl+kl_{-}}e^{-ky}+l_{-}e^{kl}+\frac{2}{k}e^{kl}-\frac{1}{k}
& ,y\in \lbrack L_{2},L_{3}]
\end{array}
\right. \ 
\end{equation}
\begin{equation}
g(z)=\left\{ 
\begin{array}{cl}
{kz+1} & ,z\in \lbrack 0,z_{1}] \\ 
{kz_{1}+1} & ,z\in \lbrack z_{1},z_{2}] \\ 
{k(z_{2}-z)+kz_{1}+1} & ,z\in \lbrack z_{2},z_{3}]
\end{array}
\right. \ 
\end{equation}
where $z_{1}=z(L_{1})$, $z_{2}=z(L_{2})$ and $z_{3}=z(L_{3})$.

The potential (\ref{pot2}) again gives rise to a massless graviton zero mode
whose wavefunction is given by (\ref{zerowave}) with the same normalization
convention. Note, however, that in the limit $l_- \rightarrow \infty$ this
mode becomes non-normalizable (GRS case). The solution of the
Schr\"{o}dinger equation for the massive KK modes is: 
\begin{equation}
\hat{\Psi}^{(n)}\left\{ 
\begin{array}{c}
\mathbf{A} \\ 
\mathbf{B} \\ 
\mathbf{C}
\end{array}
\right\}=N_n \renewcommand{\arraystretch}{1.7} \left\{ 
\begin{array}{c}
\sqrt{\frac{g(z)}{k}}\left[\phantom{A_1}Y_{2}\left(\frac{m_n}{k}%
g(z)\right)+A_{\phantom{2}}J_{2}\left(\frac{m_n}{k}g(z)\right)\right] \\ 
~~~~~~~~B_1\cos(m_{n}z)+B_{2}\sin(m_{n}z) \\ 
\sqrt{\frac{g(z)}{k}}\left[C_{1}Y_{2}\left(\frac{m_n}{k}g(z)%
\right)+C_{2}J_{2}\left(\frac{m_n}{k}g(z)\right)\right]
\end{array}
\right\} \   \label{wavemulti}
\end{equation}
where $\mathbf{A}=[0,z_{1}]$, $\mathbf{B}=[z_{1},z_{2}]$, and $\mathbf{C}%
=[z_{2},z_{3}]$. We observe that the solution in the first and third
interval has the same form as in the $^{\prime\prime}+ - +^{\prime\prime}$
model. The new feature is the second region (flat spacetime). The
coefficients that appear in the solution are determined by imposing the
boundary conditions and normalizing the wavefunction.

The boundary conditions (two for the continuity of the wavefunction at $%
z_{1} $, $z_{2}$ and four for the discontinuity of its first derivative at $%
0 $, $z_{1}$, $z_{2}$ and $z_{3}$) result in a $6\times 6$ homogeneous
linear system which, in order to have a non-trivial solution, should have
vanishing determinant. It is readily reduced to a $4\times 4$ set of
equations leading to the quantization condition: 
\begin{equation}
\renewcommand{\arraystretch}{2}{\footnotesize {\left| 
\begin{array}{cccc}
Y_{2}\left( g_{1}\frac{m}{k}\right) -\frac{Y_{1}\left( \frac{m}{k}\right) }{%
J_{1}\left( \frac{m}{k}\right) }J_{2}\left( g_{1}\frac{m}{k}\right) & -\cos
(mz_{1}) & -\sin (mz_{1}) & 0 \\ 
Y_{1}\left( g_{1}\frac{m}{k}\right) -\frac{Y_{1}\left( \frac{m}{k}\right) }{%
J_{1}\left( \frac{m}{k}\right) }J_{1}\left( g_{1}\frac{m}{k}\right) & %
\phantom{-}\sin (mz_{1}) & -\cos (mz_{1}) & 0 \\ 
0 & -\sin (mz_{2}) & \phantom{-}\cos (mz_{2}) & \phantom{-}Y_{1}\left( g_{2}%
\frac{m}{k}\right) -\frac{Y_{1}\left( g_{3}\frac{m}{k}\right) }{J_{1}\left(
g_{3}\frac{m}{k}\right) }J_{1}\left( g_{2}\frac{m}{k}\right) \\ 
0 & \phantom{-}\cos (mz_{2}) & \phantom{-}\sin (mz_{2}) & -Y_{2}\left( g_{2}%
\frac{m}{k}\right) +\frac{Y_{1}\left( g_{3}\frac{m}{k}\right) }{J_{1}\left(
g_{3}\frac{m}{k}\right) }J_{2}\left( g_{2}\frac{m}{k}\right)
\end{array}
\right| =0}}  \label{det1}
\end{equation}
with $g_{1}=g(z_{1})$, $g_{2}=g(z_{2})$ and $g_{3}=g(z_{3})$. Here we have
suppressed the subscript $n$ on the masses $m_{n}$.

\subsection{The Mass Spectrum}

The above quantization condition provides the mass spectrum of the model. It
is convenient to introduce two dimensionless parameters, $x=kl$ and $%
x_{-}=kl_{-}$ (c.f. Fig.(\ref{multi})) and we work from now on with the set
of parameters $x$, $x_{-}$ and $k$. The mass spectrum depends crucially on the
distance $x_{-}$. We must recover the KMPRS spectrum in the limit $%
x_{-}\rightarrow 0$, and the GRS spectrum in the limit $x_{-}\rightarrow
\infty .$ From the quantization condition, eq(\ref{det1}) it is easy to
verify these features and show how the $^{\prime \prime }+--+^{\prime \prime
}$ spectrum smoothly interpolates between the KMPRS model and the GRS one.
It turns out that the structure of the spectrum has simple $x_{-}$ and $x$
dependence in three separate regions of the parameter space:

\subsubsection{The three-brane \textbf{{$^{\prime \prime }+-+^{\prime \prime
}$ Region}}}

For $x_{-}\lower3pt\hbox{$\, \buildrel < \over \sim \, $}1$ we find that the
mass spectrum is effectively $x_{-}$-independent given by the approximate
form: 
\begin{eqnarray}
m_{1} &=&2\sqrt{2}ke^{-2x} \\
m_{n+1} &=&\xi _{n}ke^{-x}~~~~~~n=1,2,3,\ldots  \label{bimass}
\end{eqnarray}
where $\xi _{2i+1}$ is the $(i+1)$-th root of $J_{1}(x)$ ($i=0,1,2,\ldots $)
and $\xi _{2i}$ is the $i$-th root of $J_{2}(x)$ ($i=1,2,3,\ldots $). As
expected the mass spectrum is identical to the one of the KMPRS model for
the trivial warp factor $w=1$. The first mass is manifestly singled out from
the rest of the KK tower and for large $x$ leads to the possibility of
bigravity.

\subsubsection{The \textbf{Saturation Region}}

For $1\ll x_{-}\ll e^{2x}$ we find a simple dependence on $x_{-}$ given by
the approximate analytic form: 
\begin{eqnarray}
m_{1} &=&2k\frac{e^{-2x}}{\sqrt{x_{-}}} \\
m_{n+1} &=&n\pi k\frac{e^{-x}}{x_{-}}~~~~~~n=1,2,3,\ldots
\end{eqnarray}
As $x_{-}$ increases the first mass decreases less rapidly than the other
levels.

\subsubsection{The \textbf{GRS Region}}

For $x_{-}\gg e^{2x}$ the first mass is no longer special and scales with
respect to {\textit{both}} $x$ and $x_{-}$ in the same way as the remaining
tower: 
\begin{equation}
m_{n}=n\pi k\frac{e^{-x}}{x_{-}}~~~~~~n=1,2,3,\ldots  \label{longmass}
\end{equation}
The mass splittings $\Delta m$ tend to zero as $x_{-}\rightarrow \infty $
and we obtain the GRS continuum of states

The behaviour of the spectrum is illustrated in Figure 3.

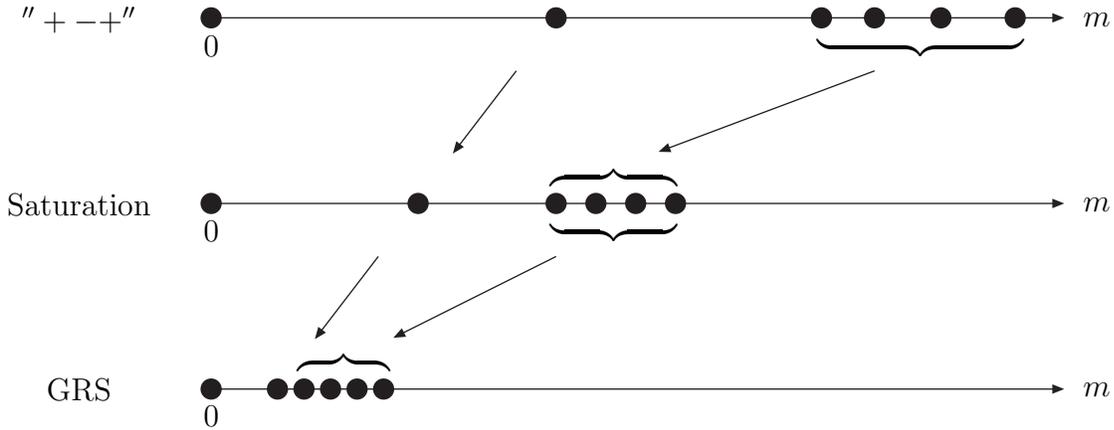
\begin{figure}[h]
\begin{center}
\begin{picture}(300,200)(0,50)

\LongArrow(0,220)(320,220)
\LongArrow(0,150)(320,150)
\LongArrow(0,80)(320,80)

\Vertex(0,220){4}
\Vertex(0,150){4}
\Vertex(0,80){4}

\Vertex(130,220){4}
\Vertex(230,220){4}
\Vertex(250,220){4}
\Vertex(275,220){4}
\Vertex(303,220){4}

\Vertex(78,150){4}
\Vertex(130,150){4}
\Vertex(145,150){4}
\Vertex(160,150){4}
\Vertex(175,150){4}

\Vertex(25,80){4}
\Vertex(35,80){4}
\Vertex(45,80){4}
\Vertex(55,80){4}
\Vertex(65,80){4}

\Text(335,220)[]{$m$}
\Text(335,150)[]{$m$}
\Text(335,80)[]{$m$}

\Text(0,210)[]{$0$}
\Text(0,140)[]{$0$}
\Text(0,70)[]{$0$}

\Text(50,85)[c]{$\overbrace{\phantom{abcdef}}$}
\Text(152,145)[c]{$\underbrace{\phantom{abcdefgh}}$}
\Text(152,155)[c]{$\overbrace{\phantom{abcdefgh}}$}
\Text(268,215)[c]{$\underbrace{\phantom{abcdefghijklm}}$}

\LongArrow(250,200)(170,170)
\LongArrow(130,130)(70,100)

\LongArrow(115,200)(92,170)
\LongArrow(63,130)(40,100)

\Text(-50,220)[c]{$''+-+''$}
\Text(-50,150)[c]{Saturation}
\Text(-50,80)[c]{GRS}

\end{picture}
\end{center}
\caption{The behaviour of the mass of the first five KK states in the three
regions of simple $x$, $x_{-}$ dependence. The first dot at zero stands for
the graviton.}
\label{masses}
\end{figure}

\subsection{Multigravity}

Armed with the details how the spectrum smoothly changes between the KPMRS
model ($x_{-}=0)$ and the GRS model ($x_{-}\rightarrow \infty )$, we can now
discuss the possibilities for modifying gravity at large distances. The
couplings of the KK states with matter on the left $^{\prime \prime
}+^{\prime \prime }$ brane are readily calculated by the interaction
Lagrangian (\ref{coupl}) with: 
\begin{equation}
a_{n}=\left[ \frac{g(0)}{M}\right] ^{3/2}\hat{\Psi}^{(n)}(0)
\label{couplmulti}
\end{equation}

\subsubsection{\textbf{Bigravity Region}}

In the KPMRS limit, $x_{-}\rightarrow 0,$ the first KK mode has constant
coupling equal to that of the 4D graviton: 
\begin{equation}
a_{1}=\frac{1}{M_{\ast }}~(=a_{0})~~~~~~~\mathrm{where}~~~M_{\ast }^{2}=%
\frac{2M^{3}}{k}
\end{equation}
while the couplings of the rest of the KK tower are exponentially
suppressed: 
\begin{equation}
a_{n+1}=\frac{1}{M_{\ast }}~\frac{e^{-x}}{\sqrt{J_{1}^{2}\left( \frac{%
m_{n}e^{x}}{k}\right) +J_{2}^{2}\left( \frac{m_{n}e^{x}}{k}\right) }}%
~~~~~~n=1,2,3,\ldots  \label{bigcoupl}
\end{equation}

The gravitational potential is computed by the tree level exchange diagrams
of the 4D graviton and KK states which in the Newtonian limit is: 
\begin{equation}
V(r)=-\sum_{n=0}^{N_{\Lambda}}a_n^2\frac{e^{-m_{n}r}}{r}  \label{gravipot}
\end{equation}
where $a_n$ is the coupling (\ref{couplmulti}) and $n=0$ accounts for the
massless graviton. The summation stops at some very high level $N_{\Lambda}$
with mass of the order of the cutoff scale $\sim M$.

In the ``bigravity'' scenario, at distances $r\ll m_{1}^{-1},$ the first KK
state and the 4D graviton contribute equally to the gravitational force, $%
i.e.$ 
\begin{equation}
V_{ld}(r)\approx -\frac{1}{M_{\ast }^{2}}\left( \frac{1}{r}+\frac{e^{-m_{1}r}%
}{r}\right) \approx -\frac{G_{N}}{r}
\end{equation}
where $G_{N}\equiv \frac{2}{M_{\ast }^{2}}$. For distances $r\gtrsim
m_{1}^{-1}$ the Yukawa suppression effectively reduces gravity to half its
strength. Astronomical constraints and the requirement of the observability
of this effect demand that for $k\sim M_{\mathrm{Pl}}$ we should have $x$ in
the region 65-70. Moreover, at distances $r\lesssim m_{2}^{-1}$ the Yukawa
interactions of the remaining KK states are significant and will give rise
to a short distance correction. This can be evaluated by using the
asymptotic expression of the Bessel functions in (\ref{bigcoupl}) since we
are dealing with large $x$ and summing over a very dense spectrum, giving: 
\begin{equation}
V_{sd}(r)=-\frac{G_{N}}{k}\sum_{n=2}^{N_{\Lambda }}\frac{k\pi }{2e^{x}}~%
\frac{m_{n}}{2k}~\frac{e^{-m_{n}r}}{r}  \label{shortpot}
\end{equation}
At this point we exploit the fact that the spectrum is nearly continuum
above $m_{2}$ and turn the sum to an integral with the first factor in (\ref
{shortpot}) being the integration measure, {\textit{i.e.}} $\sum \frac{k\pi 
}{2e^{x}}=\sum \Delta m\rightarrow \int dm$ (this follows from eq(\ref
{bimass}) for the asymptotic values of the Bessel roots). Moreover, we can
extend the integration to infinity because, due to the exponential
suppression of the integrand, the integral saturates very quickly and thus
the integration over the region of very large masses is irrelevant. The
resulting potential is now: 
\begin{equation}
V_{sd}(r)=-\frac{G_{N}}{k}\int_{m_{2}}^{\infty }dm~\frac{m}{2k}~\frac{%
e^{-m_{n}r}}{r}
\end{equation}
The integration is easily performed and gives: 
\begin{equation}
V_{sd}(r)\simeq -\frac{G_{N}}{2r}~\frac{1+m_{2}r}{(kr)^{2}}~e^{-m_{2}r}
\end{equation}
We see these short distance corrections are significant only at Planck scale
lengths $\sim k^{-1}$.

\subsubsection{The \textbf{GRS Region}}

In the GRS limit, $x_{-}\gg e^{2x},$ we should reproduce the
``resonance''-like behaviour of the coupling in the GRS model. In the
following we shall see that indeed this is the case and we will calculate
the first order correction to the GRS potential for the case $x_{-}$ is
large but finite.

For the rest of the section we split the wavefunction (\ref{wavemulti}) in
two parts, namely the normalization $N_{n}$ and the unnormalized
wavefunction $\tilde{\Psi}^{(n)}(z)$, {\textit{i.e.}} $\hat{\Psi}%
^{(n)}(z)=N_{n}\tilde{\Psi}^{(n)}(z)$. The former is as usual chosen so that
we get a canonically normalized Pauli-Fierz Lagrangian for the 4D KK modes $%
h_{\mu \nu }^{(n)}$ and is given by: 
\begin{equation}
N_{n}^{2}=\frac{1/2}{2\displaystyle{\int_{\phantom{.}0}^{z_{1}}dz\left[ 
\tilde{\Psi}^{(n)}(z)\right] ^{2}}+\displaystyle{\int_{\phantom{.}%
z_{1}}^{z_{2}}dz\left[ \tilde{\Psi}^{(n)}(z)\right] ^{2}}}
\end{equation}
The value of $\tilde{\Psi}^{(n)}(z)$ on the left $^{\prime \prime }+^{\prime
\prime }$ brane is, for $m_{n}\ll k$: 
\begin{equation}
\tilde{\Psi}_{(n)}^{2}(0)\simeq \frac{16k^{3}}{\pi ^{2}m_{n}^{4}}  \label{un}
\end{equation}

It is convenient to split the gravitational potential given by the relation (%
\ref{gravipot}) into two parts: 
\begin{equation}
V(r)=-\frac{1}{M^{3}}\sum_{n=1}^{N_{x_{-}}-1}\frac{e^{-m_{n}r}}{r}N_{n}^{2}%
\tilde{\Psi}_{(n)}^{2}(0)-\frac{1}{M^{3}}\sum_{n=N_{x_{-}}}^{N_{\Lambda }}%
\frac{e^{-m_{n}r}}{r}N_{n}^{2}\tilde{\Psi}_{(n)}^{2}(0)  \label{gravpot}
\end{equation}
As we shall see this separation is useful because the first $N_{x_{-}}$
states give rise to the long distance gravitational potential $V_{ld}$ while
the remaining ones will only contribute to the short distance corrections $%
V_{sd}$.

\begin{itemize}
\item  \textbf{\ Short Distance Corrections}
\end{itemize}

We first consider the second term. The normalization constant in this region
is computed by considering the asymptotic expansions of the Bessel functions
with argument $\frac{g(z_{1})m_{n}}{k}$. It is easily calculated to be: 
\begin{equation}
N_{n}^{2}=\frac{\pi ^{3}m_{n}^{5}}{32k^{3}g(z_{1})x_{-}}~\left[ \frac{1}{1+%
\frac{2}{x_{-}}}\right]
\end{equation}

If we combine the above normalization with unnormalized wavefunction (\ref
{un}), we find 
\begin{equation}
V_{sd}(r)\simeq -\frac{1}{M^{3}}\sum_{n=N_{x_{-}}}^{N_{\Lambda }}\frac{k\pi 
}{x_{-}e^{x}}~\frac{m_{n}}{2k}~\frac{e^{-m_{n}r}}{r}~\left[ \frac{1}{1+\frac{%
2}{x_{-}}}\right]  \label{shortpot}
\end{equation}
Since we are taking $x_{-}\gg e^{2x}$, the spectrum tends to continuum, {%
\textit{i.e.}} $N_{n}\rightarrow N(m)$, $\tilde{\Psi}_{(n)}(0)\rightarrow 
\tilde{\Psi}(m)$, and the sum turns to an integral where the first factor in
(\ref{shortpot}) is the integration measure, {\textit{i.e.}} $\sum \frac{%
k\pi }{x_{-}e^{x}}=\sum \Delta m\rightarrow \int dm$ ($c.f.$ eq(\ref
{longmass})). Moreover, as before we can again extend the integration to
infinity. Finally, we expand the fraction involving $x_{-}$ keeping the
first term in the power series to obtain the potential: 
\begin{equation}
V_{sd}(r)\simeq -\frac{1}{M^{3}}\int_{m_{0}}^{\infty }dm~\frac{e^{-mr}}{r}~%
\frac{m}{2k}\left( 1-\frac{2}{x_{-}}\right)
\end{equation}
where $m_{0}=ke^{-x}$. The integral is easily calculated and the potential
reads: 
\begin{equation}
V_{sd}(r)\simeq -\frac{G_{N}}{2r}~\frac{1+m_{0}r}{(kr)^{2}}~(1-\frac{2}{x_{-}%
})~e^{-m_{0}r}
\end{equation}
where we identified $G_{N}\equiv \frac{k}{M^{3}}$ for reasons to be seen
later. The second part of the above potential is the first correction coming
from the fact that $x_{-}$ is finite. Obviously this correction vanishes
when $x_{-}\rightarrow \infty $. Note that the above potential gives
corrections to the Newton's law only at distances comparable to the Planck
length scale.

\begin{itemize}
\item  \textbf{\ Multigravity: 4D and 5D gravity}
\end{itemize}

We turn now to the more interesting first summation in eq(\ref{gravpot}) in
order to show that the coupling indeed has the ``resonance''-like behaviour
for $\Delta m\rightarrow 0$ responsible for 4D Newtonian gravity at
intermediate distances and the 5D gravitational law for cosmological
distances. This summation includes the KK states from the graviton zero mode
up to the $N_{x_{-}}$-th level. The normalization constant in this region is
computed by considering the series expansion of all the Bessel functions
involved. It is easily calculated to be: 
\begin{equation}
N_{n}^{2}\simeq \frac{\pi ^{2}m_{n}^{4}}{4g(z_{1})^{4}x_{-}}~\left[ \frac{1}{%
m_{n}^{2}+\frac{\Gamma ^{2}}{4}+\frac{8k^{2}}{g(z_{1})^{4}x_{-}}}\right]
\end{equation}
where $\Gamma =4ke^{-3x}$. If we combine the above normalization with the
unnormalized wavefunction (\ref{un}), we find that the long distance
gravitational potential is: 
\begin{equation}
V_{ld}(r)=-\frac{1}{M^{3}}\sum_{n=0}^{N_{x_{-}}}\frac{\pi k}{x_{-}e^{x}}~%
\frac{4k^{2}}{\pi g(z_{1})^{3}}~\frac{e^{-m_{n}r}}{r}~\left[ \frac{1}{%
m_{n}^{2}+\frac{\Gamma ^{2}}{4}+\frac{8k^{2}}{g(z_{1})^{4}x_{-}}}\right]
\end{equation}
Again, since we are taking $x_{-}\gg e^{2x},$ the above sum will turn to an
integral with $\sum \Delta m\rightarrow \int dm$. Moreover, we can safely
extend the integration to infinity since the integral saturates very fast
for $m\lesssim \Gamma /4\equiv r_{c}^{-1}\ll ke^{x}$. If we also expand the
fraction in brackets keeping the first term in the power series, we find the
potential: 
\begin{eqnarray}
V_{ld}(r)\simeq &-&\frac{1}{M^{3}}\int_{0}^{\infty }dm~\frac{4k^{2}}{\pi
g(z_{1})^{3}}~\frac{e^{-mr}}{r}~\frac{1}{m^{2}+\frac{\Gamma ^{2}}{4}} 
\nonumber \\
&+&\frac{1}{M^{3}}\int_{0}^{\infty }dm~\frac{32k^{4}}{x_{-}\pi g(z_{1})^{7}}~%
\frac{e^{-mr}}{r}~\frac{1}{(m^{2}+\frac{\Gamma ^{2}}{4})^{2}}
\end{eqnarray}

The first part is the same as in the GRS model potential, whereas the second
one is the first correction that comes from the fact that $x_{-}$ is still
finite though very large. Note that the width of the ``resonance'' scales
like $e^{-3x}$, something that is compatible with the scaling law of the
masses ($m_{n}=n\pi k\frac{e^{-x}}{x_{-}}$), since we are working at the
region where $x_{-}\gg e^{2x}$, {\textit{i.e.}} $m_{n}\ll n\pi ke^{-3x}$.
The above integrals can be easily calculated in two interesting limits.

For $k^{-1}\ll r\ll r_{c}$ the potential is given approximately by : 
\begin{equation}
V_{ld}(r\ll r_{c})\simeq -\frac{G_{N}}{r}~(1-\frac{e^{2x}}{x_{-}})
\end{equation}
where we have identified $G_{N}\equiv \frac{k}{M^{3}}$ to obtain the normal
4D Newtonian potential. Note that since $x_{-}\gg e^{2x}$, the $1/x_{-}$
term is indeed a small correction.

In the other limit, $r\gg r_{c}$, the integrand is only significant for
values of $\ m$ for which the $m^{2}$ term in the denominator of the
``Breit-Wigner'' can be dropped and the potential becomes: 
\begin{equation}
V_{ld}(r\gg r_{c})\simeq -\frac{G_{N}r_{c}}{\pi r^{2}}~(1-\frac{2e^{2x}}{%
x_{-}})
\end{equation}
The fact that Newtonian gravity has been tested close to the present horizon
size require that for $k\sim M_{\mathrm{Pl}}$ we should have $x\gtrsim $
45-50.

Finally we note that if we take the $x_{-}\rightarrow \infty $ we recover
the GRS result 
\begin{equation}
\lim_{x_{-}\rightarrow \infty }V_{+--+}(r,x_{-})=V_{GRS}(r)
\end{equation}

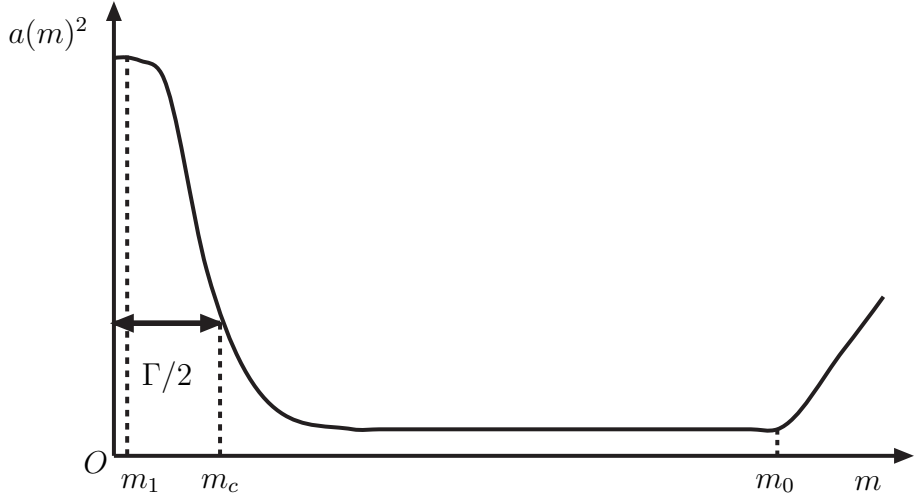
\begin{figure}[tbp]
\begin{center}
\begin{picture}(400,200)(-50,50)
\SetOffset(50,0)
\SetWidth{1.5}

\LongArrow(-50,80)(250,80)
\LongArrow(-50,80)(-50,250)
\Curve{(-50,230)(-40,229)(-32,224)(-15,150)(40,90)(50,90)(75,90)(100,90)(125,90)(150,90)(175,90)(190,90)(200,90)(225,120)(240,140)}
\DashLine(-45,80)(-45,231){2.5}
\DashLine(-10,80)(-10,130){2.5}
\DashLine(200,80)(200,90){2.5}

\Text(-90,240)[l]{$a(m)^2$}
\Text(-40,70)[c]{$m_{1}$}
\Text(-30,110)[c]{$\Gamma/2$}
\Text(230,70)[l]{$m$}
\Text(200,70)[c]{$m_{0}$}
\Text(-10,70)[c]{$m_{c}$}
\Text(-57,78)[c]{$O$}

\SetWidth{2}
\LongArrow(-50,130)(-10,130)
\LongArrow(-10,130)(-50,130)

\end{picture}
\end{center}
\caption{The behaviour of the coupling, $a(m),$ in the limit of $x_{-}\gg
e^{2x}$. Three regions of interest are indicated. The region $m>m_{0}$ gives
rise to short distance corrections. The $m_{1}\ll m\ll m_{c}$ region gives
rise to 4D gravity at intermediate distances and 5D gravity at ultra large
distances. For distances $r\gg m_{1}^{-1}$, the zero mode gives the dominant
contribution and thus we return to 4D gravity.}
\label{coupling}
\end{figure}

\begin{itemize}
\item  \textbf{\ Back to 4D gravity}
\end{itemize}

As we have just seen, probing larger distances than $r_{c},$ the 4D
gravitational potential changes to a 5D one. This is the most significant
characteristic of the GRS model. In the case that $x_{-}$ is large compared
to $e^{2x}$ but still finite, there is another distinct region of interest,
namely $r\gg m_{1}^{-1}$. This follows from the fact that, in this limit,
the spectrum is still discrete. For distances larger than of the order of
the corresponding wavelength of the first KK mode, the contribution to
gravity from the KK tower is suppressed and thus the zero mode gives the
dominant contribution, leading to the 4D Newtonian potential again. In this
case the strength of the gravitational interaction is a small fraction of
the strength of the intermediate 4D gravity. More precisely, the
contribution of the massless graviton is $\frac{1}{x_{-}}$ suppressed and
thus vanishes when $x_{-}\rightarrow \infty $, something that is expected
since in this limit there is no nomalizable zero mode. The gravitational
potential in this case is: 
\begin{equation}
V_{4D}(r)=-\frac{1}{M^{3}}~\frac{1}{r}~N_{0}^{2}\tilde{\Psi}_{(0)}^{2}(0)=-%
\frac{G_{N}}{r}~\frac{e^{2x}}{x_{-}}
\end{equation}
Obviously this 4D region disappears at the limit $x_{-}\rightarrow \infty $
since the spectrum becomes continuum and thus the 5D gravity ``window''
extents to infinity. We should note finally that for the values of $x$ that
we consider here this final modification of gravity occurs at distances well
above the present horizon.

\section{The Crystal Universe Model}

The last model we consider is the crystal universe introduced in \cite
{kaloper} (see also \cite{nam}). It consists of an
infinite array of parallel 3-branes in a 5D $AdS$ space with cosmological
constant $\Lambda $. For simplicity we assume that the branes are
equidistant with distance, $l$, between two successive branes. Needless to
say, in this case all the $^{\prime \prime }+^{\prime \prime }$ branes have
warp factor $w=1$ with respect to a $^{\prime \prime }+^{\prime \prime }$
brane sitting at the origin of the 5-th dimension coordinate. The metric
ansatz that has 4D Poincar\'{e} invariance is again given by eq(\ref{ansatz}%
) where the $\sigma (y)$ function is constrained by the the Einstein
equations to have the sawtooth form: 
\begin{equation}
\sigma (y)=k\sum_{j=-\infty }^{+\infty }(y-2jl)\left[ 2\theta (y-2jl)-\theta
(y-(2j-1)l)-\theta (y-(2j+1)l)\right]
\end{equation}
The tensions of successive branes are required to be opposite and equal to $%
\pm \Lambda /k$, where $k=\sqrt{\frac{-\Lambda }{24M^{3}}}$ is a measure of
the curvature of the bulk and $M$ the 5D fundamental scale.

\begin{figure}[t]
\begin{center}
\begin{picture}(300,80)(0,50)
\SetOffset(0,-20)
\SetWidth{1.5}

\Line(-40,100)(340,100)
\GCirc(150,100){7}{0.9}
\GCirc(10,100){7}{0.9}
\GCirc(290,100){7}{0.9}

\GCirc(80,100){7}{0.2}
\GCirc(220,100){7}{0.2}

\LongArrow(160,110)(210,110)
\LongArrow(210,110)(160,110)
\Curve{(185,110)(180,130)(170,140)}
\Text(168,140)[r]{$x=kl$}

\Text(80,75)[c]{$-$}
\Text(220,75)[c]{$-$}
\Text(150,75)[c]{$+$}
\Text(10,75)[c]{$+$}
\Text(290,75)[c]{$+$}

\Text(80,125)[c]{$-l$}
\Text(220,125)[c]{$l$}
\Text(10,125)[c]{$-2l$}
\Text(290,125)[c]{$2l$}

\Text(340,125)[c]{$\cdots$}
\Text(-40,125)[c]{$\cdots$}

\end{picture}
\end{center}
\caption{The Crystal Universe made up of an infinite array of $^{\prime
\prime }+^{\prime \prime }$ and $^{\prime \prime }-^{\prime \prime }$ branes
with lattice \mbox{
spacing $l$} and bulk curvature $k$.}
\label{crystal}
\end{figure}
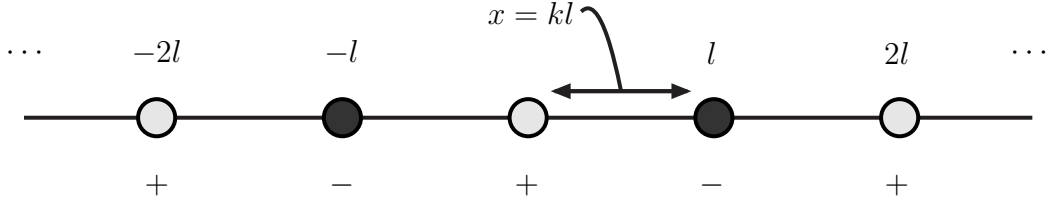

We consider the general fluctuations around the previous vacuum ansatz of
the form (\ref{fluct}) and decompose as usual the perturbation tensor
$h_{\mu \nu }(x,y)$ as: 
\begin{equation}
h_{\mu \nu }(x,y)=\int dm\,h_{\mu \nu }(m,x)\Psi (m,y)
\end{equation}
where $\Psi (m,y)$ is a complex function that arises from the linear
combination of the independent graviton polarizations. This leads to the
Schr\"{o}dinger equation (\ref{sch}) for the wave function $\hat{\Psi}(m,z)\equiv \Psi (m,y)e^{\sigma /2}$ with potential: 
\begin{equation}
V(z)=\frac{15k^{2}}{8[g(z)]^{2}}-\frac{3k}{2g(z)}\sum_{j=-\infty }^{+\infty
}(-)^{j}\delta (z-z_{i})
\end{equation}

The $z$ coordinates have been defined as usual to be the ones that make the
background metric conformally flat. In these coordinates the branes sit at
the points \mbox{$z_j=j\frac{e^{kl}-1}{k}\equiv jz_l$} and the function $%
g(z) $ is: 
\begin{equation}
g(z)=1+k\sum_{j=-\infty }^{+\infty }(z-z_{2j})\left[ 2\theta
(z-z_{2j})-\theta (z-z_{2j-1})-\theta (z-z_{2j+1})\right]
\end{equation}

The solution of the Schr\"{o}dinger equation for the wavefunctions for two
adjacent cells is given\cite{kaloper} in terms of Hankel functions: 
\begin{equation}
\hat{\Psi}(m,z)=N_{m}\sqrt{\frac{g(z)}{k}}\left[ \left\{ 
\begin{array}{c}
1 \\ 
B \\ 
e^{2iqz_{l}} \\ 
e^{2iqz_{l}}B
\end{array}
\right\} H_{2}^{+}\left( \frac{m}{k}g(z)\right) +\left\{ 
\begin{array}{c}
A \\ 
C \\ 
e^{2iqz_{l}}A \\ 
e^{2iqz_{l}}C
\end{array}
\right\} H_{2}^{-}\left( \frac{m}{k}g(z)\right) \right]  \label{sol}
\end{equation}
in the regions $z\in \lbrack 0,z_{l}]$, $z\in \lbrack z_{l},2z_{l}]$, $z\in
\lbrack 2z_{l},3z_{l}]$ and $z\in \lbrack 3z_{l},4z_{l}]$ respectively 
\footnote{%
We use as \cite{kaloper} Hankel instead of real Bessel functions in order to
encode the gravitons phase rotation.}. In the above expression $N_{m}$ is an
overall normalization constant, $q$ is the Bloch wave quasi-momentum and the
constants $A$,$B$,$C$ are given by: 
\begin{eqnarray}
A &=&\frac{(h_{1}^{+}h_{2}^{-}+h_{1}^{-}h_{2}^{+})\hat{h}%
_{2}^{+}-(h_{1}^{-}h_{2}^{+}-h_{1}^{+}h_{2}^{-})e^{2iqz_{l}}\hat{h}%
_{2}^{+}-2h_{1}^{+}h_{2}^{+}\hat{h}_{2}^{-}}{%
(h_{1}^{+}h_{2}^{-}+h_{1}^{-}h_{2}^{+})\hat{h}%
_{2}^{-}+(h_{1}^{-}h_{2}^{+}-h_{1}^{+}h_{2}^{-})e^{2iqz_{l}}\hat{h}%
_{2}^{-}-2h_{1}^{-}h_{2}^{-}\hat{h}_{2}^{+}} \\
B &=&\frac{h_{1}^{+}h_{2}^{-}+h_{1}^{-}h_{2}^{+}+2h_{1}^{-}h_{2}^{-}A}{%
h_{1}^{-}h_{2}^{+}-h_{1}^{+}h_{2}^{-}} \\
C &=&\frac{2h_{1}^{+}h_{2}^{+}+(h_{1}^{+}h_{2}^{-}+h_{1}^{-}h_{2}^{+})A}{%
h_{1}^{+}h_{2}^{-}-h_{1}^{-}h_{2}^{+}}
\end{eqnarray}
where $h_{n}^{\pm }\equiv H_{n}^{\pm }\left( \frac{m}{k}g(z_{l})\right) $
and $\hat{h}_{n}^{\pm }\equiv H_{n}^{\pm }\left( \frac{m}{k}\right) $. The
above coefficients of the Hankel functions were determined by the boundary
conditions, {\textit{i.e.}} continuity of the wavefunction at $z_{l}$, $2z_{l}$, $%
3z_{l} $, discontinuity of its first derivative at $z_{l}$, $3z_{l}$ and the
Bloch wave conditions relating the wavefunction at the edges of each cell.
The last remaining boundary condition of the discontinuity of the first
derivative of the wavefunction at $2z_{l}$ gives us the band equation
connecting $q$ and $m$. We disagree with the relation given in \cite{kaloper}
and instead we find: 
\begin{equation}
\cos(2qz_{l})=\frac{(j_{2}y_{1}+j_{1}y_{2})(\hat{\jmath}_{2}\hat{y}_{1}+\hat{%
\jmath}_{1}\hat{y}_{2})-2\hat{\jmath}_{1}\hat{\jmath}%
_{2}y_{1}y_{2}-2j_{1}j_{2}\hat{y}_{1}\hat{y}_{2}}{(j_{2}y_{1}-j_{1}y_{2})(%
\hat{\jmath}_{2}\hat{y}_{1}-\hat{\jmath}_{1}\hat{y}_{2})}\equiv f(m)
\label{band}
\end{equation}
where again $j_{n}\equiv J_{n}\left( \frac{m}{k}g(z_{l})\right) $, $%
y_{n}\equiv Y_{n}\left( \frac{m}{k}g(z_{l})\right) $ and $\hat{\jmath}%
_{n}\equiv J_{n}\left( \frac{m}{k}\right) $, $\hat{y}_{n}\equiv Y_{n}\left( 
\frac{m}{k}\right) $. This dispersion relation {\textit{always}} gives a
band at zero as is to be expected intuitively. Defining the parameter $%
x\equiv kl$, we see that for $x\lower3pt%
\hbox{$\, \buildrel > \over \sim \,
$}10$ we may reliably approximate the width of the zero mode band by: 
\begin{equation}
\Gamma _{0}=2\sqrt{2}ke^{-2x}
\end{equation}
while the separation of the zeroth and the first band is 
\begin{equation}
\Delta \Gamma _{1}=\xi _{1}ke^{-x}
\end{equation}
with $\xi _{1}$ the 1st root of $J_{1}(x)$. This characteristic behaviour of
the above widths with respect to $x$ implies that we can have a viable
multigravity scenario if $\Gamma _{0}^{-1}\lower3pt%
\hbox{$\, \buildrel >
\over \sim \, $}10^{26}\mathrm{cm}$ and at the same time $\Delta \Gamma
_{1}^{-1}\lower3pt\hbox{$\, \buildrel < \over \sim \, $}1\mathrm{mm}$. This
can happen for $k\sim M_{\mathrm{Pl}}$ and $x\approx 68$.

The band structure has the following form. The first bands have very narrow
widths $\Gamma _{i}$ with spacing $\Delta \Gamma _{i}$ between them ({%
\textit{i.e.}} between the $i$-th and the $(i-1)$-th band), which are
approximately: 
\begin{eqnarray}
\Gamma _{i} &\approx &\epsilon _{i}\xi _{i}^{2}ke^{-3x} \\
\Delta \Gamma _{i} &\approx &(\xi _{i}-\xi _{i-1})ke^{-x}
\end{eqnarray}
where $\xi _{2n+1}$ is the $(n+1)$-th root of $J_{1}(x)$, $\xi _{2n}$ is the
{}-th root of $J_{2}(x)$, and 
\mbox {
$\epsilon_{2n+1}=\frac{\pi}{4}
\frac{Y_1(\xi_{2n+1})}{J_2(\xi_{2n+1})}$}, $\epsilon _{2n}=\frac{\pi }{4}%
\frac{Y_{2}(\xi _{2n})}{J_{1}(\xi _{2n})}$. As we move on to higher bands
their width is increasing and the spacing between them is decreasing. In the
limit $m\rightarrow \infty $, the spacing disappears, {\textit{i.e.}} $\Delta \Gamma
_{i}\rightarrow 0$, while their width tends to $\Gamma _{i}\rightarrow \frac{%
\pi }{2}ke^{-x}$. In this limit the function $f(m)$ in (\ref{band}) tends to 
$f(m)=\cos(2z_{l}m)$ and we have no forbidden zones.

In order to find the gravitational potential we should at first find how the
KK modes couple to matter and determine their spectral density. Instead of
working with wavefunctions in an infinite extra dimension with continuous
normalization, it is more convenient effectively to compactify the system to
a circle of length $2{\mathcal {N}}l$, by assuming that we have a finite crystal of ${\mathcal {N}}$
cells $[2jl,2(j+1)l]$. We shall find the gravitational potential is
independent of ${\mathcal {N}}$, so we will be able to take the limit ${\mathcal {N}}\rightarrow
\infty $ and decompactify the geometry.

In this regularization scheme we can readily see that the ever-present zero
mode, corresponding to the 4D graviton, has the wavefunction: 
\begin{equation}
\hat{\Psi}(0,z)=\frac{\sqrt{k/{\mathcal {N}}}}{[g(z)]^{3/2}}  \label{grav}
\end{equation}
and its coupling to matter on the central positive brane is: 
\begin{equation}
a(0)=\frac{1}{\sqrt{\mathcal {N}}}\sqrt{\frac{k}{M^{3}}}\equiv \frac{1}{\sqrt{\mathcal {N}}}\frac{1}{%
M_{\ast }}
\end{equation}

The wavefunction eq(\ref{grav}) is a smooth limit of eq(\ref{sol}) for $m\rightarrow 0$. Indeed, for large $x$ we find the normalization constant in
(\ref{sol}) in the region of the zero mode band is approximately: 
\begin{equation}
N_{m}=\frac{\sqrt{2}}{z_{l}^{2}m}\frac{1}{\sqrt{\mathcal {N}}}
\end{equation}
The approximation of the Hankel functions with their first term of their
power series proves our previous claim. Furthermore, we can see that the KK
states in the whole zero mode band for large $x$ will couple to matter in
the central positive brane as: 
\begin{equation}
a(m)=\frac{1}{\sqrt{\mathcal {N}}}\frac{1}{M_{\ast }}e^{i\phi (m)}
\end{equation}
with $\phi (m)$ an unimportant m-dependent phase that doesn't appear in
observable quantities.

In the limit $m\rightarrow \infty $ the Hankel functions can be approximated
by their asymptotic form. The wavefunction is then almost constant in
z-space and the KK states couple to matter on the central positive brane as: 
\begin{equation}
a(m)=\frac{1}{\sqrt{\mathcal {N}}}\frac{1}{\sqrt{2e^{x}}M_{\ast }}e^{i\phi (m)}
\end{equation}
where again $\phi (m)$ is an unimportant m-dependent phase. The couplings in
the low lying bands interpolate between the above limiting cases.

The standard procedure to calculate the spectral density is at first to find
the number of states with masses smaller than $m$ and then differentiate
with respect to $m$. In our regularization prescription it is
straightforward to count the states in the quasi-momentum space with masses
lighter that $m$. The latter is simply: 
\begin{equation}
N_{tot}(m)=\frac{{\mathcal {N}}z_l}{\pi}q
\end{equation}
so the spectral density is its derivative: 
\begin{equation}
\rho(m)=\frac{dN_{tot}(m)}{dm}=\frac{{\mathcal {N}}z_l}{\pi} \frac{dq}{dm}= (\mp) \frac{{\mathcal {N}}}{2 \pi} \frac{1}{\sqrt{1-f(m)^2}} \frac{df(m)}{dm}
\end{equation}
where the plus and minus signs will succeed each other for neighboring
bands, starting with minus for the zero mode band. For the zero band we can
have a reliable approximation for large $x$ of the above function: 
\begin{equation}
\rho(m)={\mathcal {N}} \frac{1}{\pi \Gamma_0} \frac{1}{\sqrt{1 -
\left(m/\Gamma_0\right)^2}}
\end{equation}

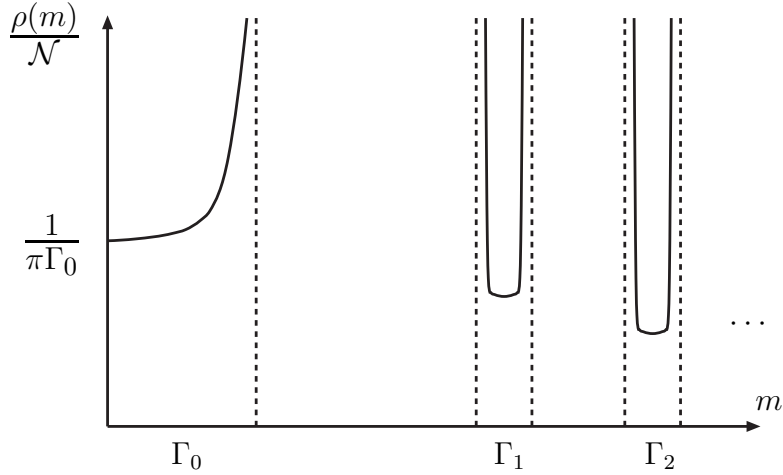
\begin{figure}[t]
\begin{center}
\begin{picture}(300,160)(0,40)
\SetScale{0.7}
\SetOffset(70,20)
\SetWidth{1.5}

\LongArrow(-50,30)(300,30)
\LongArrow(-50,30)(-50,250)
\DashLine(30,30)(30,250){3}

\Text(-45,170)[r]{\Large {$\frac{\rho(m)}{\mathcal {N}}$}}
\Text(-45,90)[r]{\Large {$\frac{1}{\pi \Gamma_0}$}}
\Text(210,30)[l]{$m$}
\Text(200,60)[l]{$\cdots$}

\Curve{(-50,130)(-35,131)(-15,134)(-8,136)(0,141)(5,146)(15,175)(25,250)}

\Text(-5,10)[c]{$\Gamma_0$}
\Text(117,10)[c]{$\Gamma_1$}
\Text(174,10)[c]{$\Gamma_2$}

\SetOffset(90,20)
\DashLine(120,30)(120,250){3}
\DashLine(150,30)(150,250){3}
\DashLine(200,30)(200,250){3}
\DashLine(230,30)(230,250){3}
\Curve{(125,250)(126,130)(126.5,110)(127,104)(128,102)(130,101)(135,100)(140,101)(142,102)(143,104)(143.5,110)(144,130)(145,250)}
\Curve{(205,250)(206,110)(206.5,90)(207,84)(208,82)(210,81)(215,80)(220,81)(222,82)(223,84)(223.5,90)(224,110)(225,250)}

\end{picture}
\end{center}
\caption{Spectral density for the first three bands}
\label{density}
\end{figure}

Thus the spectral density diverges as $m \rightarrow \Gamma_0$ but the
divergence is integrable and doesn't cause any problem to the following
calculations. Actually, the spectral density diverges at the edges of every
band since at these points $\sqrt{1-f(m)^2} \rightarrow 0$ while $\frac{df(m)%
}{dm} \neq 0$. The only point that this doesn't happen is at $m=0$ where $%
\frac{df(m)}{dm} \rightarrow 0$ in such a way that the result is finite.

In the limit $m\rightarrow \infty $ the bands disappear and the spectrum has
constant spectral density: 
\begin{equation}
\rho (m)={\mathcal {N}}\frac{e^{x}}{k\pi }
\end{equation}

The gravitational potential, taking into account the Yukawa suppressions of
the KK states, is simply: 
\begin{equation}
V(r)=-\int_{0}^{\infty }dm|a(m)|^{2}\rho (m)\frac{e^{-mr}}{r}
\end{equation}

For distances $r\gg \Delta \Gamma _{1}^{-1}$ we have the effective
potential: 
\begin{eqnarray}
V(r) &=&-\frac{1}{M_{* }^{2}}\frac{1}{r}\int_{0}^{\Gamma _{0}}dm\frac{
\rho (m)}{{\mathcal {N}}}e^{-mr}=-\frac{1}{\pi M_{\ast }^{2}}\frac{1}{r}\int_{0}^{1}d\xi 
\frac{e^{-(\Gamma _{0}r)\xi }}{\sqrt{1-\xi ^{2}}}  \nonumber \\
&=&-\frac{1}{2M_{\ast }^{2}}\frac{1}{r}[I_{0}(\Gamma _{0}r)-L_{0}(\Gamma
_{0}r)]
\end{eqnarray}
where $I_{0}$ is the $0$-th modified Bessel function and $L_{0}$ is the $0$
-th modified Struve function. The gravitational potential is ${\mathcal {N} }$
-independent as expected. For distances $\Delta \Gamma _{1}^{-1}\ll r\ll
\Gamma _{0}^{-1}$ the above functions tend to $I_{0}(\Gamma
_{0}r)\rightarrow 1$ and $L_{0}(\Gamma _{0}r)\rightarrow 0$, so we recover
the 4D Newton law with $G_{N}=\frac{1}{2M_{\ast }^{2}}$: 
\begin{equation}
V(r\ll \Gamma _{0}^{-1})=-\frac{G_{N}}{r}
\end{equation}

On the other hand, for distances $r \gg \Gamma_0^{-1}$ the asymptotic
expansion of the difference of the modified Bessel and Struve functions
gives a 5D Newton law: 
\begin{equation}
V(r \gg \Gamma_0^{-1})= -\frac{2G_N}{\pi \Gamma_0} \frac{1}{r^2}
\end{equation}

In case we take the crystal to be finite, there will appear a region for $%
r\gg m_{1}^{-1}$ where gravity will turn again to 4D. However, as explained
in the previous section, this region is well above the universe horizon and
thus of no phenomenological interest.

Finally, for distances $r\lower3pt\hbox{$\, \buildrel < \over \sim \, $}%
\Delta \Gamma _{1}^{-1}$ we will start to feel the short distance Yukawa
type modifications to gravity due to the presence of the bands above the
zeroth one. As in the zero hierarchy ``bigravity'' and $^{\prime \prime
}+--+^{\prime \prime }$ ``multigravity'' model we expect these corrections
to be important at scales of order the Planck length.

\section{Discussion and Conclusions}

We have shown that in models in which negative tension branes are sandwiched
between positive tension branes there are anomalously light KK states that
can lead to modifications of gravity. So far we have implicitly assumed that
the gravitational force generated by a massive graviton is identical to that
generated by a massless one but this is not, in general, the case because
the massive graviton has additional degrees of freedom which do not decouple
in the massless limit. As a result the interaction generated by a massive
graviton violates the normal 4D relation between the gravitational
interactions of matter and light. This may be seen explicitly from the form
of the massive and massless graviton propagators. Up to terms involving
momentum vectors which do not contribute to
$T_{\mu\nu}G^{\mu \nu , \alpha \beta}T_{\alpha \beta}$ due to momentum 
conservation, the propagator has the form \cite{VZ,KK} 
\begin{equation}
G^{\mu \nu ,\alpha \beta }(x-x^{\prime })=\int \frac{dp^{4}}{(2\pi )^{4}}%
\frac{1/2\left( g^{\mu \alpha }g^{\nu \beta }+g^{\nu \alpha }g^{\mu \beta
}\right) -tg^{\mu \nu }g^{\alpha \beta }}{p^{2}-m^{2}-i\epsilon }%
e^{-ip(x-x^{\prime })}
\end{equation}
where for $m\neq 0$, $t=1/3$ but for $m=0$, $t=1/2$. The
difference between the two propagators 
\[
\delta G^{\mu \nu ,\alpha \beta }(x-x^{\prime })=\frac{1}{6}\int \frac{dp^{4}%
}{(2\pi )^{4}}\frac{g^{\mu \nu }g^{\alpha \beta }}{p^{2}-m^{2}-i\epsilon }%
e^{-ip(x-x^{\prime })} 
\]
is due to the additional helicity components needed for a massive graviton.
There are three additional components needed. Two, corresponding to the
graviphotons, decouple at low energies as they are derivatively coupled to
the conserved energy-momentum tensor. The third, corresponding to a scalar
component, does not decouple and is responsible for $\delta G.$ The
contribution of $\delta G$ to the one-graviton exchange amplitude for any
two four-dimensional sources $T_{\mu \nu }^{1}$ and $T_{\alpha \beta }^{2}$
vanishes for light for which $T=T_{\mu }^{\mu }=0$ but not for matter
because $T=T_{0}^{0}\neq 0$, hence the normal 4D gravity relation between
the interactions of matter and light is violated as was first pointed
out by Dvali et al \cite{dvali1}. This is inconsistent with observation, for example the bending of light by the sun.

It has been argued that this violation does \textit{not} occur in the models
discussed above. Two mechanisms have been identified. The first \cite{rest} involves a cancellation of the contribution from $\delta G$ due
to an additional scalar field, the radion, the moduli field associated with
the size of the new dimension. This cancellation was related to the
bending of the brane due to the matter sources following the approach
suggested by Garriga and Tanaka \cite{garr}. This field must have unusual properties
because it is known that a normal scalar contributes with the same sign as $%
\delta G$ and therefore cannot cancel it. An explicit computation of Pilo et
al \cite{pilo} of the properties of the radion in the GRS model show
that it indeed has a negative kinetic energy term leading to a contribution
which cancels the troublesome $\delta G$ contribution. As a result the
theory indeed gives rise to 4D gravity at intermediate distances. The
presence of the ghost radion field is worrying for the consistency of the
theory for it gives rise to ``antigravity'' at extremely large scales when
the theory become five dimensional. Moreover as the ghost energy is
unbounded from below it is likely that any theory is unstable when coupled
to such a state \cite{dvali2}.
It has been suggested that the problems associated with the
ghost apply to all theories in which there is a negative tension brane
sandwiched between positive tension branes because such theories violate the
weak energy condition \cite{CEH,witten}. The latter has been shown \cite{warner}
to require $\sigma ^{\prime \prime }(y)\geq 0$ and, as we have seen, models
which have intermediate negative tension branes violate this condition.
However it is not clear that theories that violate the weak energy condition
are unacceptable \cite{visser}. Indeed it has been demonstrated that any
theory involving a scalar field necessarily violates the condition \cite
{visser1} and the condition also does not apply to the gravitational sector 
\cite{mavromatos}. Certainly more work is needed to clarify these issues.

The second mechanism capable of cancelling the $\delta G$ contribution was
identified in \cite{KR} offers some hope that the problems associated with
the ghost state may be avoided. The mechanism follows from the underlying
5D structure of the theory. The graviton propagator in five dimensions has
the tensor structure \cite{KK} 
\begin{equation}
G^{MN,PQ}\sim \left[ 1/2\left( g^{MP}g^{NQ}+g^{MQ}g^{NP}\right)
-tg^{MN}g^{PQ}\right] +O(pp/p^{2})
\end{equation}
where $t=1/3$ i.e.the same as for the \textit{massive} four-dimensional
graviton. The full five-dimensional one graviton exchange amplitude has the
form 
\begin{equation}
T_{MN}^{1}G^{MN,PQ}T_{PQ}^{2}\varpropto T_{MN}^{1}\left[ 1/2\left(
g^{MP}g^{NQ}+g^{MQ}g^{NP}\right) -tg^{MN}g^{PQ}\right] T_{PQ}^{2}
\end{equation}
where $T_{MN}$ is the five dimensional stress-energy tensor. In the
compactifications considered here $T_{\mu 5}$ vanishes but in general $%
T_{55} $ does not. Thus the total amplitude can be written as 
\[
T_{\mu \nu }^{1}T^{2\mu \nu
}-tT^{1}T^{2}+(1-t)T_{5}^{15}T_{5}^{25}-t(T_{5}^{15}T^{2}+T^{1}T_{5}^{25})= 
\left[ T_{\mu \nu }^{1}T^{2\mu \nu }-\frac{1}{2}T^{1}T^{2}\right] 
\]
\begin{equation}
+(1/2-t)T^{1}T^{2}+(1-t)T_{5}^{15}T_{5}^{25}-t(T_{5}^{15}T^{2}+T^{1}T_{5}^{25})
\label{amp}
\end{equation}
The first term in square brackets is the normal four-dimensional amplitude
corresponding to massless graviton exchange. The second term takes account
of the $\delta G$ contribution as well as the contribution from $T_{5}^{5}.$
It may be written in the form 
\begin{equation}
\frac{1}{6}(T^{1}-2T_{5}^{15})(T^{2}-2T_{5}^{25})
\end{equation}
In \cite{KR} it was pointed out that this contribution necessarily vanishes
for stable brane configurations. To see this we first note that all
observable amplitudes involving states confined to the visible sector brane
have zero momentum transfer in the fifth direction. This means that we have
to integrate over the fifth coordinate $\int dy\,\sqrt{-G^{(5)}}%
\,(T-2T_{5}^{5})$. However it was shown in \cite{kkop1} that this
combination vanishes in any theory which stabilises the extra dimension. As
a result the contribution of the second line of eq(\ref{amp}) corresponding
to $\delta G$ vanishes and one is left with the usual four dimensional
gravitational interaction. Note that this explanation does not require a
field with negative kinetic energy as it arranges for the additional
(scalar) graviton component associated with a massive graviton to decouple
from matter via a cancellation of the various contributions to the stress
energy tensor.

As discussed above there may still be a troublesome negative kinetic energy
radion field but now it too decouples because it also couples to $\int dy\,%
\sqrt{-G^{(5)}}\,(T-2T_{5}^{5}).$ This may eliminate the problems associated
\ with a negative kinetic energy scalar. Indeed it suggests that we should
be able to introduce an additional field to cancel the ghost field without
disturbing the 4D structure of the gravitational interactions.

In summary, in this paper we discussed three models based on different brane
configurations, the three-brane $^{\prime \prime }++-^{\prime \prime }$
model, the four-brane $^{\prime \prime }+--+^{\prime \prime }$
``multigravity'' model and the crystalline ``multigravity'' model. The $%
^{\prime \prime }++-^{\prime \prime }$model provides an instructive example
showing that we cannot have special light KK states (and thus
``multigravity'') without intermediate $^{\prime \prime }-^{\prime \prime }$
branes. This model, however, does provide a way of having the visible sector
on a $^{\prime \prime }+^{\prime \prime }$ brane with an hierarchical
``warp'' factor. The second $^{\prime \prime }+--+^{\prime \prime }$
``multigravity'' model provides an interpolation between the KMPRS and the
GRS models and leads to a new possibility in which gravity changes from 4D
to 5D and then back again to 4D. The third model, the infinite crystalline
``multigravity'' model provides a further example of how the
``multigravity'' scenario appears every time that we have $^{\prime \prime
}-^{\prime \prime }$ branes between $^{\prime \prime }+^{\prime \prime }$
branes. Again gravity changes from 4D to 5D at ultralarge distances (and
back to 4D if the crystal is finite). While models with intermediate
negative tension branes have very interesting phenomenological implications
there is a question whether they are consistent as they necessarily violate
the weak energy condition \cite{witten,warner}. Related to this is the fact
that they may have ghost states with negative kinetic energy. We have shown
that if the brane configurations are stabilised these ghost states decouple,
offering the possibility that the problems associated with such states may
be avoided. Moreover the same decoupling condition ensures that the
gravitational interactions associated with the KK excitations of gravity
responsible for ``multigravity'' couple in the usual 4D way to matter and
radiation.

\textbf{Acknowledgments:} We would like to thank N. Mavromatos for
usefull discussions. S.M.'s work is supported by the Hellenic State
Scholarship Foundation (IKY) \mbox{No. 
8117781027}. A.P.'s work is supported by the Hellenic State Scholarship
Foundation (IKY) \mbox{No. 8017711802}. The work of I.I.K. and G.G.R. is
supported in part by PPARC rolling grant PPA/G/O/1998/00567, the EC TMR
grant FMRX-CT-96-0090 and by the INTAS grant RFBR - 950567.

\end{document}